\documentclass[aps,prb,longbibliography,reprint]{revtex4-2}

\usepackage[english]{babel}
\usepackage[utf8]{inputenc}
\usepackage[colorinlistoftodos, color=green!40, prependcaption]{todonotes}
\usepackage{graphicx}
\usepackage{bm}

\usepackage{amsmath}

\usepackage{xcolor}

\begin{document}

\title{Radiation of Optical Angular Momentum by a Dipole Source inside a Magneto-optical Environment }

\author{B.A. van Tiggelen}
\email[]{Bart.Van-Tiggelen@lpmmc.cnrs.fr}
\affiliation{Univ. Grenoble Alpes, CNRS, LPMMC, 38000 Grenoble, France}

\date{\today} 

\begin{abstract}
Radiation of electromagnetic energy by electric or magnetic multipole sources can be  modified by their local environment. In this work we demonstrate that a magneto-optical environment
of an unpolarized dipole source induces the radiation of angular momentum into space. This radiation benefits from Purcell-enhancement.
\end{abstract}

\keywords{}

\maketitle

\section{Introduction}
The power emitted by a light source  depends on its local environment.
This notion, strongly related to the well-know Purcell effect \cite{purcell},  can be quantified by the local density of
electromagnetic modes available to the photons leaving the source. It has important applications, e.g. to enhance weak magnetic dipole transitions \cite{magnetoPurcell} or to suppress or enhance spontaneous emission \cite{jablo, JMG, pbgPurcell}. The statement holds for classical light sources and for quantum sources only if they have low quantum efficiency \cite{dary}.
Landau and Lifshitz \cite{LL}  calculate the torque on a classical emitting dipole in vacuum as an exercise and find zero. A magneto-birefringent environment changes this outcome and induces the radiation of angular momentum into space \cite{pinwheel}.
In this work we calculate how much angular momentum is
radiated into space by a classical dipole inside a magneto-active Mie sphere that acts as a (resonant) nano-antenna \cite{nanoantennapurcell} for angular momentum .
 By conservation of angular momentum a torque is exerted on the source, proportional to the power of the source.

\section{Radiation of energy and optical angular momentum}

The radiation of electromagnetic waves by a classical source is described  by the vector Helmholtz equation for the electric field (in Gaussian units),

\begin{equation}
- \frac{   \varepsilon(\bf{r})}{c_0^2}\cdot \partial_t^2 {\bf E}({\bf{r}},t) -{\bf\nabla}\times {\bf \nabla}\times {\bf E}({ \bf r},t)  = \frac{4\pi}{c_0^2} \partial_t {\bf J}_S({\bf r},t)
\end{equation}\label{HH0}
 featuring the current density $\bf{J}_S$ of the classical source. Its direct environment is characterized by (here only) the dielectric tensor
$\varepsilon_{ij}(\bf{r})$. We shall ignore its frequency dependence, neglect absorption and give up explicit reference to frequency dependence. For a monochromatic source at frequency $\omega$ , the solution can be formally written
in Dirac notation as ${\bf E} ({\bf r})= \langle {\bf r}| {\bf E}\rangle$ with   $|{\bf E}\rangle  ={\bf G} \cdot (-4\pi i\omega /c_0^2)|{\bf J}_S \rangle $. The Helmholtz Green's operator given by

\begin{equation}\label{HH}
{ \bf G}({\bf x},{\bf p}) =    \frac{1}{ \varepsilon ({\bf x})(k_0+i0)^2 - p^2 + {\bf pp}}
\end{equation}
where $k_0=\omega/c_0$ and $ {\bf x}, {\bf p}$ the usual canonical operators, obeying $\left[ {x}_n, {p}_m\right] =i\delta_{nm}$. The small positive imaginary part ensures the outwardly propagating solution relevant for a source. According to Maxwell's equations the cycle-averaged dissipation  density is
$ \frac{1}{2} \mathrm{Re}  \left({\bf J}_S^*\cdot {\bf E} \right) $. The emitted the power is therfore,
\begin{equation}\label{PS0}
P_S =  - \frac{2\pi k_0}{c_0} \mathrm{Im} \int d^3{\bf r} \int d^3{\bf r}' \,  {\bf J}_S({\bf r})^*\cdot {\bf G}({\bf r},{\bf r}')\cdot {\bf J}_S({\bf r}')
\end{equation}
For a source with electric dipole (ED) moment ${\bf d}$ is  ${\bf J}_{\mathrm{ED}}({\bf r}) = -i\omega {\bf d} \delta({\bf r})$, for a magnetic dipole (MD) moment ${\bf m}$ is
${\bf J}_{\mathrm{MD}}({\bf r}) = c_0 {\bf \nabla} \times \left[{\bf m} \delta({\bf r}) \right]$. It follows \cite{magnetoPurcell}

\begin{eqnarray}\label{PS}
P_{\mathrm{ED}} &=& -\frac{2\pi}{3} k_0^3c_0 |{\bf d}|^2 \mathrm{Im} \, G_{ii}(0,0)  \nonumber \\
P_{\mathrm{MD}} &=& -\frac{2\pi}{3} k_0c_0 |{\bf m}|^2 \mathrm{Im} \, \hat{G}_{ii}(0,0)
\end{eqnarray}
with ${\hat{G}}_{kk'} = (i\epsilon_{knl}p_l) G_{nn'} (i\epsilon_{n'k'l'}p_{l'})$ the Helmholtz vector Green's function associated with the magnetic field.  The same can be done for the radiation
 of angular momentum by the source at $r\approx 0$. The electromagnetic angular momentum is defined as $ {\bf J}_{\rm rad} =\mathrm{ Re} \int d^3 {\bf r}\,  {\bf r}\times ({\bf E}^* \times  {\bf B})/8\pi c_0$ that can be split up into an orbital, a spin and a longitudinal
 component \cite{cc, barAM}.
 Without external torques the sum of ${\bf J}_{\rm rad}$ and mechanical angular momentum ${\bf J}_{\rm mec}$ of the source and its environment must be conserved.
 Let us consider a radius $R$ enclosing the environment.
 For a source that has been stationary during a time $t\gg R/c_0$, total electromagnetic angular momentum inside the sphere $r < R$ should be steady. The balance becomes \cite{LL}
\begin{eqnarray}
0 &=& -{ M}_i + \frac{R^3}{8\pi}  \epsilon_{ijk} \mathrm{Re }\int_{4\pi} d\hat{{\bf r} } \, \hat{r}_l \hat{r}_j(E_l^*E_k + B_l^*B_k)(R\hat{{\bf r} } ) \nonumber \\
\frac{d}{dt} {J}_{i, \mathrm{mec}}  &=& +{M}_i
\end{eqnarray}
This expresses that the total torque ${\bf M}$ on the matter is radiated away as angular momentum to infinity ($r>R)$. The torque acts on both source and its environment, ${\bf M} = {\bf M}_S + {\bf M}_E$. The latter is
\begin{equation}\label{ME}
{\bf M}_E = \frac{1}{2} \mathrm{Re} \, \int d^3{\bf r} \, \left[ {\bf P}^* \times  {\bf E} + P_m^* ({\bf r} \times  {\bf \nabla}) E_m \right]
\end{equation}
but will  vanish for a rotationally-invariant environment. The torque on the source does not, and has an electric and magnetic part,
 \begin{equation}
{\bf M}_S = \frac{1}{2} \mathrm{Re} \, \int d^3{\bf r} \, \left[{\rho}_S^*({\bf r} \times  {\bf E}  ) + {\bf r} \times \left( \frac{1}{c_0}{\bf J}_S^*  \times {\bf B}  \right)\right]
\end{equation}
For an ED or MD this simplifies to expressions similar to Eqs.~(\ref{PS}),
\begin{eqnarray}\label{MSS}
{M}_{i, \mathrm{ED}} &=& -\frac{2\pi}{3} k_0^2 |{\bf d}|^2 \mathrm{Re}\, \epsilon_{ijk} { G}_{kj}(0,0)  \nonumber \\
M_{i, \mathrm{MD}} &=& - \frac{2\pi}{3}  |{\bf m}|^2 \mathrm{Re} \, \epsilon_{ijk} { \hat{G}}_{kj}(0,0)
\end{eqnarray}
These torques vanish for an environment with isotropic optical response \cite{LL}. In the presence of magneto-birefringence,
the torque is directed along the external magnetic field ${\bf B}_0 $. Both vectors  have same parity but behave differently under time-reversal, yet the relation
${\bf M}_S  \sim {\bf B}_0 $ is allowed  because the time-reversed process corresponds to the \emph{absorption }of an
incident spherical wave
by a
sink, implying  $i0 \rightarrow -i0$ in Eq.~(\ref{HH}).  This changes the sign of $P_S$, whereas ${\bf M}_S$ is left unchanged
because  ${\bf B}_0$ changes sign as well upon time-reversal.
The power $P_S$ is emitted as a radial Poynting vector
that decays as $S_r= P_S /4\pi r^2$ in the far field. The radiated angular momentum emerges as a vortex with an azimuthal Poynting vector $S_\phi$ around ${\bf B}_0 $
that decays with distance as $-M_S c_0 \sin \theta / 4\pi r^3 $.
Hence the radiation of angular momentum is associated with
a small ``photon Hall angle" $S_\phi/S_r = -(M_S c_0/P_S)\sin\theta /r$ that slowly decays in the far field \cite{pinwheel}.

\section{Application to a Mie sphere}
We model the local environment of the source by a homogeneous dielectric sphere subject to  magneto-optical birefringence surrounding it symmetrically. For an external magnetic field ${\bf B}_0$ in the $z$-direction, the dielectric tensor of the sphere is
given by $\varepsilon = m^2 {\bf 1} - W(r){\bf S}_z$ with ${S}_{z,ij}= -i\epsilon_{ijz}$ the electromagnetic spin operator along the magnetic field\cite{cc}. The dimensionless material parameter $W(r)= 2mV(r)B_0/k_0$ involves
the Verdet constant $V(r)$ that may vary inside the sphere $r < a$ or even be outside. It is small, and we treat the birefringence as a linear perturbation to the Mie sphere. If   ${\bf G}^M$ is the Helmholtz vector Green's function for a Mie sphere with index of refraction $m$,
we can write Eq.~(\ref{HH}) as
   \begin{equation}\label{Gex}
{\bf G} = {\bf G}^M + {\bf G}^M \cdot k_0^2 W(r){\bf  S}_z \cdot {\bf G} ^M   + \mathcal{O}(W^2)
\end{equation}
We expand ${\bf G}^M$ into the orthonormal set of Mie eigenfunctions, involving  the spherical vector harmonics $\hat{{\bf Z}}_{JM}, \hat{{\bf N}}_{JM}, \hat{{\bf X}}_{JM}$ that are
also eigenfunctions of  total electromagnetic angular momentum ${\bf J}^2$ and its component along the $z$-axis (with eigenvalue $M$)\cite{cc}. To this end  we write

\begin{equation}
{ \bf G}^M=  \varepsilon^{-1/2} \cdot \frac{1}{(k_0+i0)^2 - \mathcal{H}}\cdot  \varepsilon^{-1/2}
\end{equation}
with $ \mathcal{H} = \varepsilon^{-1/2} \cdot (p^2 -{\bf  pp} ) \cdot \varepsilon^{-1/2}$   a hermitian and semi-positive-definite operator. It has positive eigenvalues $k^2$ associated with transverse magnetic ($e$) and
transverse electric ($m$) eigenfunctions, and  an infinitely degenerated subspace of eigenfunctions with eigenvalue $0$ associated with longitudinal electric fields. We can decompose it as
\begin{equation}\label{GMM}
{ \bf G}^M({\bf r},{\bf r}') = \int_0^\infty dk \sum_{i=e,m,\ell} \sum_{J=1}^\infty \sum_{M=-J}^J \frac{{\bf E}^i_{JMk}({\bf r}) {\bf E}^i_{JMk}({\bf r}')^* }{(k_0+i0)^2-k_i^2}
\end{equation}
with $k_{e,m}=k$ and $k_\ell=0$ for all $k$. The longitudinal mode with $J=0$ is ignored since it carries no angular momentum.
For magneto-birefringence inside the sphere we need the fields modes inside, properly normalized in the far field \cite{Mersch}. With $y \equiv mkr$ and $\tilde{y}
\equiv kr/m $,
\begin{eqnarray}
 &&{\bf E}^e_{JMk}({\bf r})  =  \frac{4\pi  k  A^{e} _{kJ} }{(2\pi)^{3/2}} \left( \sqrt{J(J+1)} \frac{j_J(y)}{y}  \hat{{\bf N}}_{JM}(\hat{{\bf r}}) \right. \nonumber \\  && \ \ \ \ \ \ \  + \left. \frac{(yj_J(y))' }{y}\hat{{\bf Z}}_{JM}(\hat{{\bf r}}) \right) \nonumber \\
 &&{\bf E}^m_{JMk}({\bf r})  =  \frac{4\pi  k  A^{m}_{kJ}}{(2\pi)^{3/2}}  j_J(y)  \hat{{\bf X}}_{JM}(\hat{{\bf r}}) \nonumber  \\
&&  {\bf E}^\ell_{JMk}({\bf r})  =  \frac{4\pi  k  A^{\ell} _{kJ}}{(2\pi)^{3/2} }   \left( j_J'(\tilde{y})  \hat{{\bf N}}_{JM}(\hat{{\bf r}}) + \sqrt{J(J+1)} \frac{j_J(\tilde{y}) }{\tilde{y}}\hat{{\bf Z}}_{JM}(\hat{{\bf r}}) \right)
\nonumber \\
\end{eqnarray}
The factors $A^{m/e}_{kJ}$ can be found in standard literature \cite{newton, vdh}, and $A^{\ell}_{kJ} = m/[ mj_J'(ka/m)h_J(ka) - j_J(ka/m) h'_J(ka) ]$ is less well known, because not excited
in the scattering of a transverse plane wave.
Upon inserting Eq.~(\ref{Gex}) and Eq.~(\ref{GMM}) into expression~(\ref{MSS}) for an ED source,   we find that only modes with $i=e,\ell$ and $J=1$ survive,
 \begin{eqnarray}\label{FED}
&& M_{\mathrm{ED},i} =   - \frac{2\pi^2 k_0^3|{\bf d}|^2}{3} \mathrm{Re} \int_0^\infty dk \sum_{i=e,\ell}^{M=\pm 1} \frac{M \langle  {\bf E}^{e}_{k_01M} | W(r) {\bf S}_z |  {\bf E}^i_{k1M}\rangle }{ k_0^2 -k_i^2 +i0}
\nonumber \\ &&  \ \ \ \ \  \  \times \ {\bf E}^e_{k_01M}(0)  \cdot {\bf E}^i_{k1M}(0)^* \nonumber \\
&& P_{\mathrm{ED}} = |A^{e} _{k_01}|^2 \times \frac{k_0^4  |{\bf d}|^2c_0}{3}
\end{eqnarray}
The matrix element is proportional to $M$ \cite{lacoste} and reduces to a radial integral over de sphere.
 For the radiating MD we find similarly,
 \begin{eqnarray}\label{FMD}
 &&M_{\mathrm{MD},i} =   \frac{2\pi^2  k_0 |{\bf m}|^2}{3} \nonumber \\ 
  &&  \ \ \ \ \  \  \times \
 \mathrm{Re} \int_0^\infty dk \sum_{M=\pm 1} \frac{M \langle  {\bf E}^m_{k_01M} | W(r) {\bf S}_z |  {\bf E}^m_{k1M}\rangle }{ k_0^2 -k^2 +i0}
 \nonumber \\ &&  \ \ \ \ \  \  \times \
 \frac{k_0^2k^2}{3\pi^2 }m^2 A^m_{1k}A^{m*}_{1k_0}  \nonumber \\
&& P_{\mathrm{MD}} = m^2|A^{m} _{k_01}|^2 \times \frac{k_0^4  |{\bf m}|^2c_0}{3}
\end{eqnarray}
where we have evaluated the electric field modes at $r=0$ explicitly.
In both cases, the $k$-integral  suffers from oscillatory behavior and slow convergence at large $k$. Some singularities result from the singular dipole field that can be extracted  using  the asymptotic expressions for
$|A^i_{kJ}|^2 \sim 2/\left[m^2+1 \pm (m^2-1)\cos (2k_{\mathrm{in}}a) \right]$ and, for $ r < 2a$,
  \begin{eqnarray}\label{ana}
&&\frac{1}{ {2\pi^2 }  r }\int_0^\infty dk \frac{ 2k j_1(kr)}{m^2+1 \pm (m^2-1)\cos(2ka)} \nonumber \\
\\ &&  \ \ \ \ \  \   \  =\frac{1}{4\pi m r^3} + \frac{2\delta({\bf r})  }{3(m^2+1)}
\end{eqnarray}
The singular dipole field has no effect on either torque or power.

\begin{figure}[t]
\includegraphics[width=5cm, angle=-90]{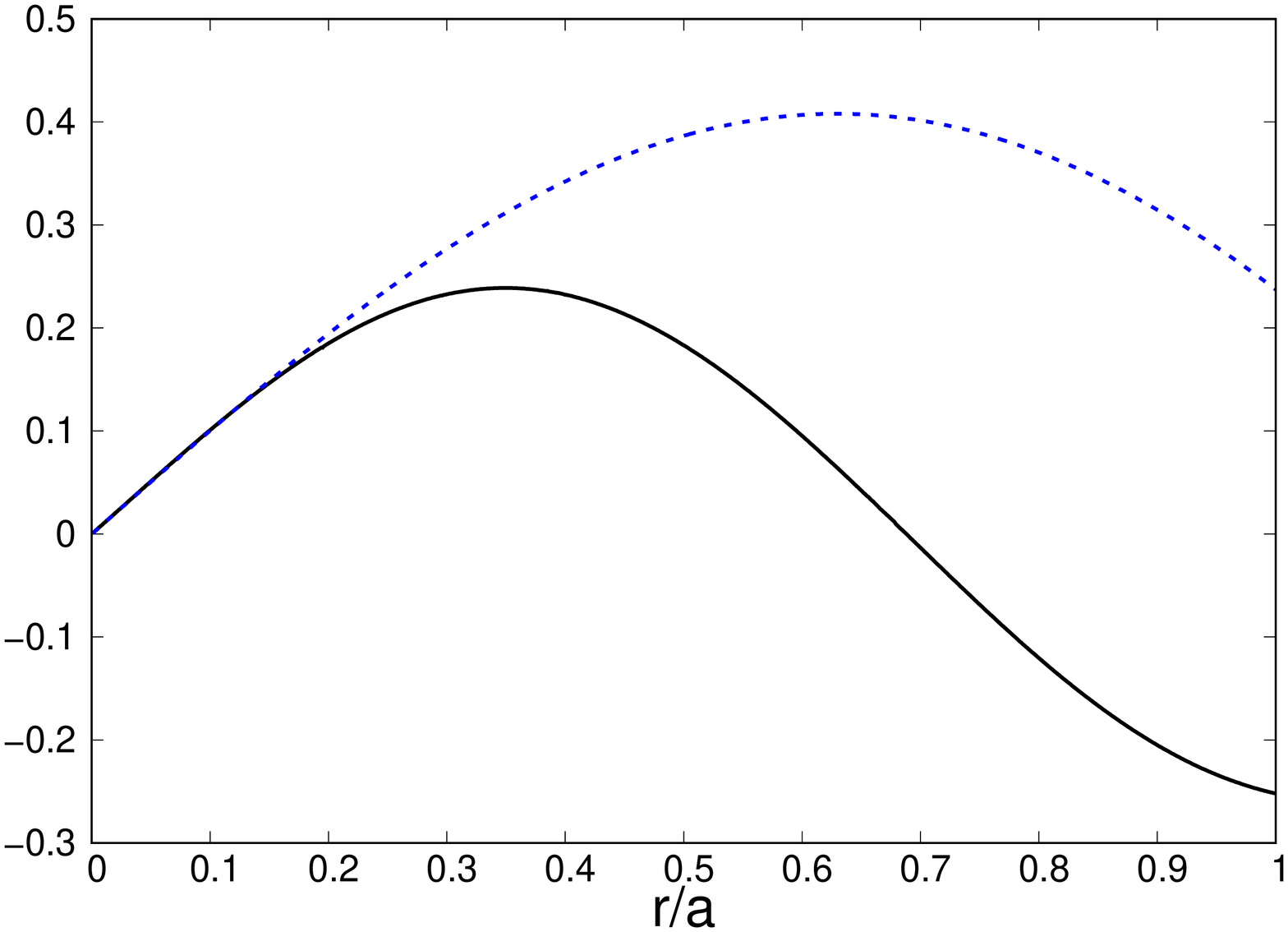}
\includegraphics[width=5cm, angle=-90]{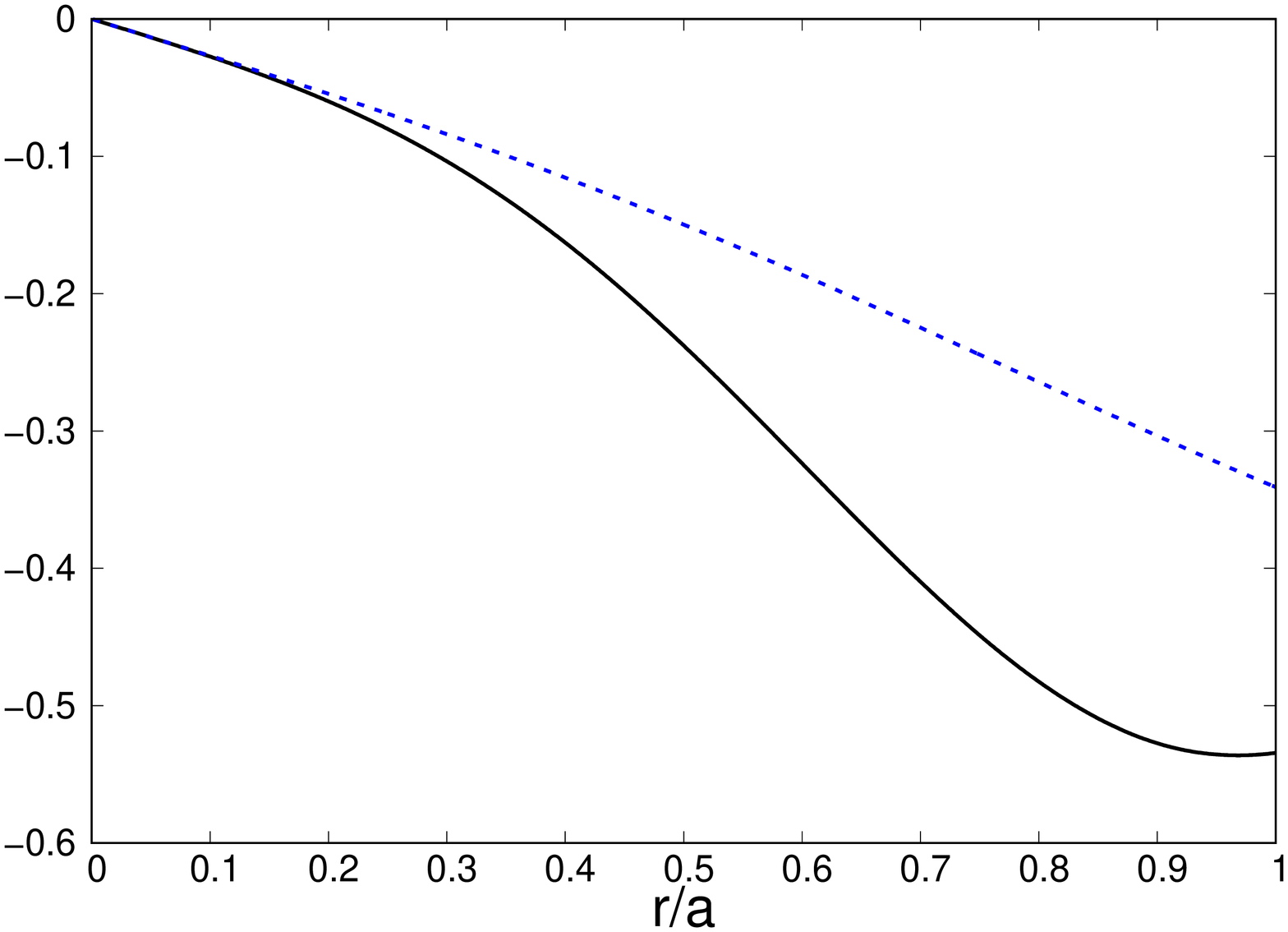}
\caption{The function $\Phi_{S}(r)$ as a function of distance inside the dielectric sphere with index of refraction $m=2$ and size parameter
$k_0a = 1.2$, for an electric dipole source (top), and a magnetic dipole source (bottom).  The blue dashed lines represent the analytical result for $m=1$. }
\label{figm2}
\end{figure}

\begin{figure}[t]
\includegraphics[width=5cm, angle=-90]{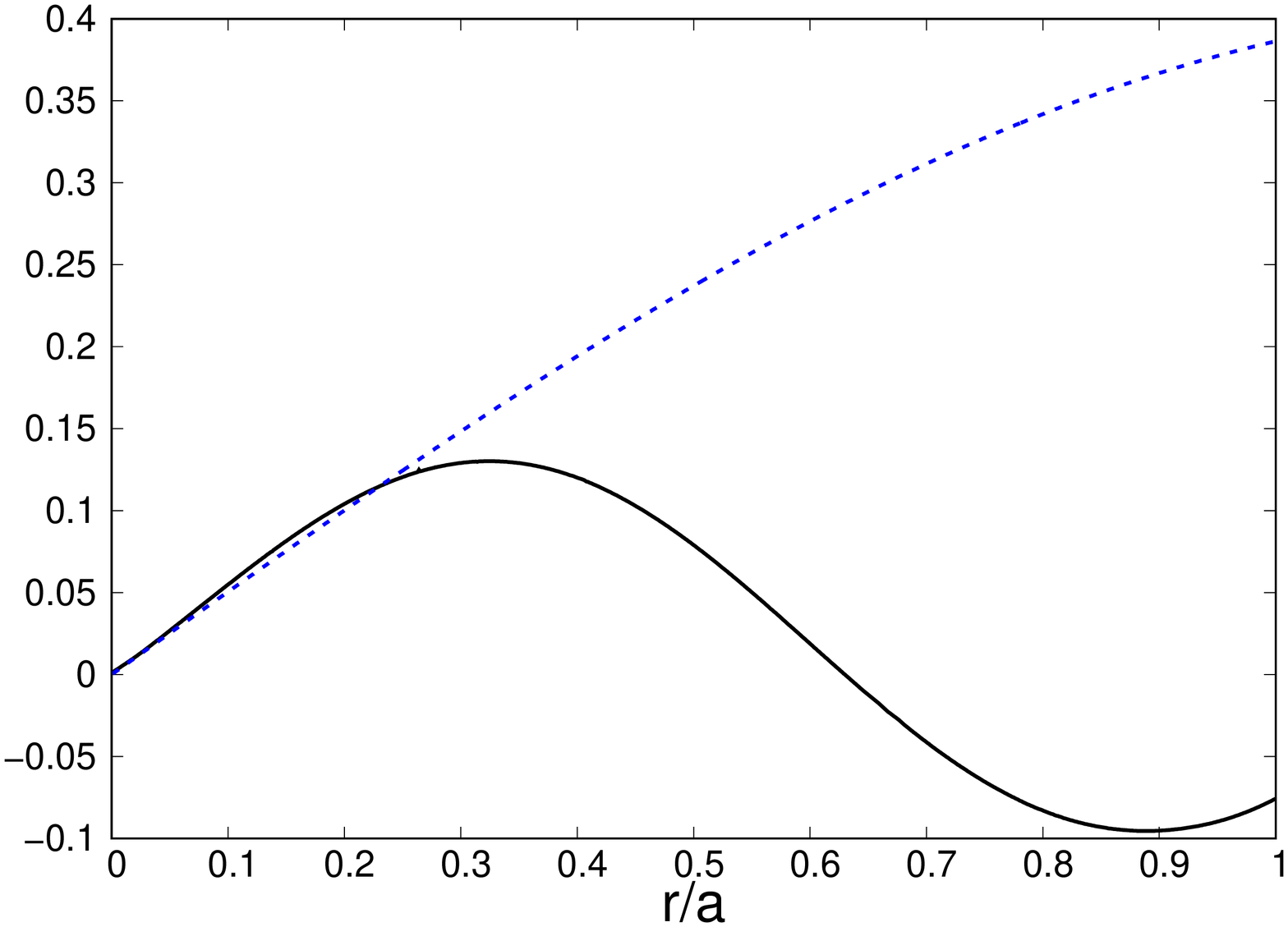}
\includegraphics[width=5cm, angle=-90]{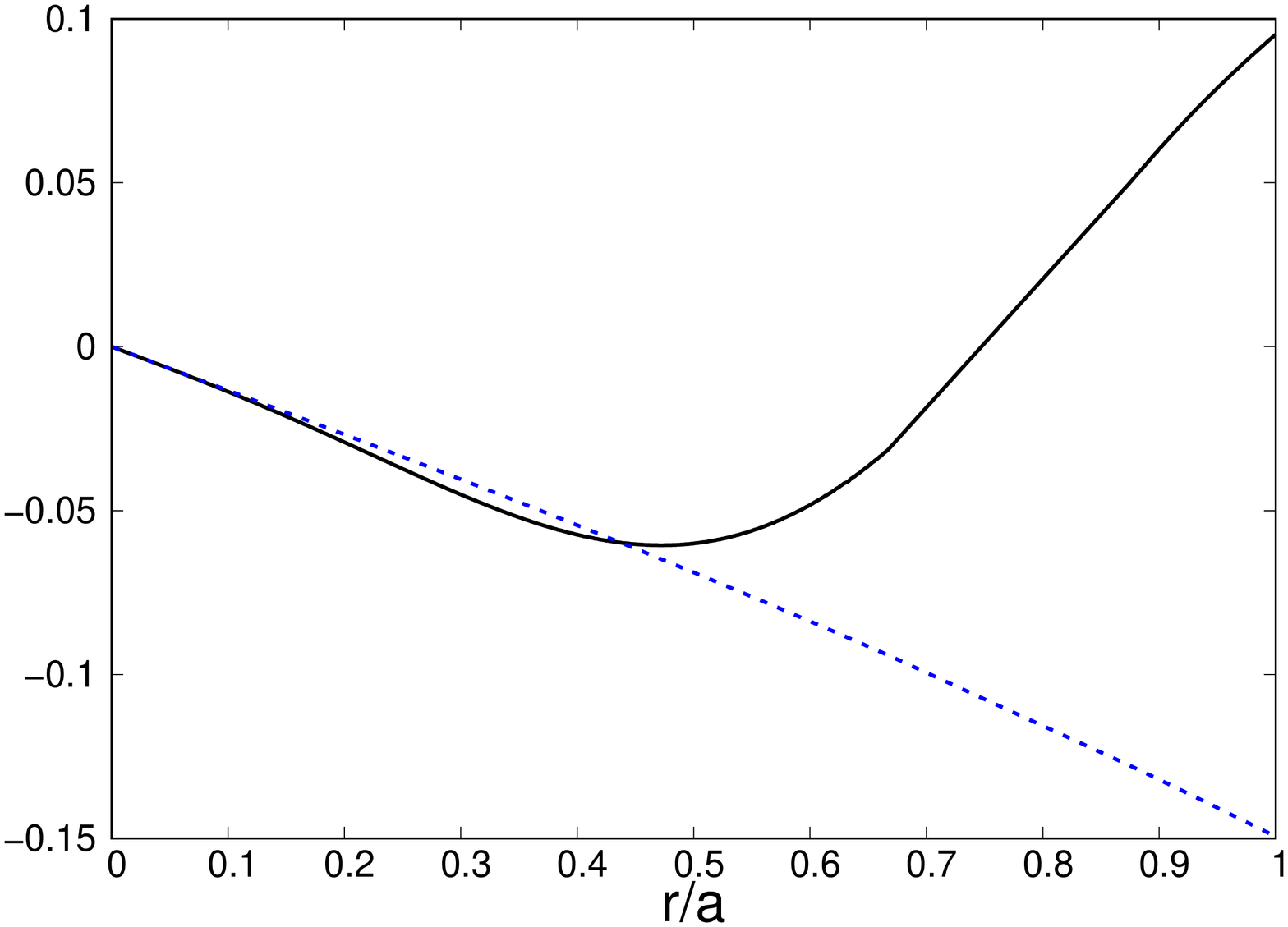}
\caption{The function $\Phi_{S}(r)$ as a function of distance inside the dielectric sphere with index of refraction $m=5.1$ and size parameter
$k_0a = 0.599$ corresponding to a magnetic dipole resonance. Top: electric dipole source; Bottom: magnetic dipole source.
The blue dashed lines represent the analytical result for $m=1$. }\label{figm51}
\end{figure}

\section{Numerical results}
We have numerically evaluated the torques in Eqs.~(\ref{FED}) and~(\ref{FMD}) and show the results in Figs.~(\ref{figm2},\ref{figm51}) for $m=2$ and $m=5.1$.
We calculate the dimensionless ratio $ M_S\omega /P_S $ that measures the amount of emitted angular momentum in units of $\hbar$ per emitted photon. The ratio $ M_S /P_S $ also
determines the photon Hall angle in the far field as mentioned above. The Purcell factors $|A^e_{1k_0}|^2$ and $m^2|A^m_{1k_0}|^2$  for ED and MD respectively cancel
in this ratio.
Because we focus here on magneto-birefringence inside de sphere  we can  write,
\begin{equation}\label{Mdr}
    \frac{M_S \omega}{P_S} =  k_0\int_0^a dr W(r) \Phi_S(r) =  2mVB_0 a \times \eta_S
\end{equation}
The second equality holds for constant $W(r)$ inside the sphere. The  entire local environment $r < a$ contributes but not always with constant sign, and singular dipole fields cancel at $r=0$. For $m=1$ we can derive analytically that
$\Phi_{\mathrm{ED}}(r)= (\sin x + 4j_1(x) -16 j_2(x)/x)/3 $ and $\Phi_{\mathrm{MD}}(r)= (\sin x - 4j_1(x))/3 $ with $x=2k_0a$.
The linear slopes of these curves in the vicinity of  the source  seem to survive for all cases $m>1$ studied. For $r/a > 0.25$ the radial profile changes significantly and can have different sign.
 From Table 1 we see that the ratio, including its sign, still depends  on index of refraction $m$ and size parameter $k_0a$. The Mie parameters are chosen for illustration and do not correspond to known magneto-optical materials. Large Purcell factors $\left|A^{e/m}_{1k_0}\right|^2$ indicate the presence of a ED or MD resonance.
For not too large $m$ there is no significant further enhancement in the ratio~(\ref{Mdr}) near resonances, because the $k$-integral in Eq.~(\ref{FED}) is less sensitive to
what happens exactly on-shell ($k=k_0$).
For large $m$  this is different. The extreme values $m=9.1$ and $k_0a = 0.341$ in Table 1 correspond to a sharp MD resonance, studied for water droplets in the microwave region  \cite{water}.
The ratio $ {M_S \omega}/{P_S} $ is much larger than for the similar MD resonance at  $m=5.1$ and $m=7.1$, and roughly follows the Purcell factor $\left|mA^m_{1k_0}\right|^2$.
Also at
the first ED resonance the ratio $M_{\mathrm{ED}} \omega /P_{\mathrm{ED}}$ is  a factor $20$  larger than at the ED resonance for $m=5$, roughly equal to the ratio of ED Purcell factors. In these extreme cases, the torque $M_{S} $ scales almost quadratically with the Purcell factor. We mention also that the longitudinal Mie eigenfunction is never negligible  to calculate the ED torque.

\begin{table}[b]
\center \begin{tabular}{|c|c||c|c|c|c| }
  \hline
$k_0a$ & $m$ & $\left|A^e_{1k_0}\right|^2$ & $\left|mA^m_{1k_0}\right|^2$ & $\eta_{ED}$  &  $ \eta_{MD}$ \\ \hline\hline
     1.0 & 1.05 & 0.987 & 1.096 & 0.25  & -0.13\\
   1.2 & 2.0 & 1.008 & 8.712 & -0.020  & -0.31\\
      1 & 1.582 & 0.849 & 1.851 & 0.046  & -0.15 \\
       1.5 & 1.582 & 1.093 & 4.867 & 0.014  &   -0.40 \\
        2 & 1.582 & 1.243 & 6.128 &  -0.0013 & -0.051\\
         3 & 1.582 & 2.157 & 2.428 &  -0.13 & -0.22 \\
          3.5 & 1.582 &  1.233 & 4.302 &  -0.25 & -0.57\\
  2.5 & 2.73 & 1.764 & 14.66 & 0.17 & 0.24\\
  4.5 & 2.73 &  1.077 &44.84 & -0.12  & -0.68 \\
  0.855 & 5.0 & 41.00 & 12.28& 0.020  & 0.0066\\
    0.599 & 5.1 & 0.171 & 2343 & 0.0096  & 0.0089\\
     0.617 & 7.1 & 282.1 & 15.32 & -0.023  & 0.0014 \\
    0.435 & 7.1 & 0.0427 & 15462 & 0.011  & -0.049 \\
        0.487 & 9.1 & 1039 & 17.10 & -0.41  & 0.00043\\
        0.341 & 9.1 & 1.508 & 55805  & 0.0024 & -0.37\\
  \hline
\end{tabular}
\caption{\label{Tmie} Numerical values for the ratio of torque to power $\eta_S \equiv (M_S\omega/P_S)/2mVB_0 a  $ for different Mie parameters, off and on resonance,
normalized to the  dimensionless material
parameter $2mVB_0 a$ in Eq.~(\ref{Mdr}) associated with Faraday rotation, and assumed constant inside the sphere.
For large index of refraction $m$ the precision is only  $ 5 \%$ due to  slowly converging
and oscillating integrals. The value $m= 1.582$ corresponds to  the paramagnetic compound EuF$_2$  that has large Verdet constant.}
\end{table}

\section{Discussion and Conclusion}
The approach above reveals that the magneto-active sphere acts as a nano-antenna for angular momentum. The emitted angular momentum feeds back to the source as a  torque and emerges as a vortex field of the Poynting vector around the environment.
This is a new effect and shows once more that the emitted radiation couples back to the source  after having interacted with its environment.
The rate of radiated angular momentum is proportional to the product of a material parameter $2mVBa$,
the  power of the source $P_S$ and the small value $\eta_S$ calculated in the two columns on the right of Table 1 that quantifies the local environment.
Because magneto-optical effects are small, the emitted angular momentum is clearly small. As an example we mention EuF$_2$ that has a huge paramagnetic Verdet constant
$V \approx -70.000$ rad/T/m at $4.5$ K and $\lambda= 466$ nm \cite{magneto} and previously used for optical magneto-resistance measurements \cite{Anja}. For $x=3.5$, $\lambda= 488 $nm implies $a= 271$ nm, about the EuF$_2$ particle size in Ref.~\cite{Anja}, so that  $2mVBa= -6.0 \cdot 10^{-2} $  at $B_0= 1$ Tesla. The calculated value $\eta_{\mathrm{ED}}=-0.25$ implies that each emitted photon sends an amount of angular momentum  $  + 0.015 \times \hbar$ along the magnetic field into space.
The radiated angular momentum cumulates in time and is subject to  resonant enhancement. Phase-sensitive detection may be used to enhance the
effect even further \cite{pinwheel}.

The nature of the emitted  angular momentum is either  spin or orbital but this has not been investigated here.
A magneto-optical coating outside the dielectric sphere might be easier to design experimentally.
Equations~(\ref{FED}) and (\ref{FMD}) remain valid and without absorption the effect does not fall with the distance of the coating to the source. Because the environment is supposed to be rotationally-invariant, no torque is exerted on the environment. For any  disorder in the environment, this will probably change. A related intriguing  question is if the Poynting vortex  will exert a radiation force on a gas bounded outside the environment of the source.

\section{Acknowledgments}
The author thanks Romuald Le Fournis for drawing attention to Ref.~\cite{LL}.

Numerical data underlying the results presented in this paper are not publicly available at this time but may be obtained from the author upon reasonable request.

  \end{document}